# Electromagnetic stress at the boundary: photon pressure or tension?

Shubo Wang,[1] Jack Ng,[2] Meng Xiao,[1] and C. T. Chan[1,*]

[1]Department of Physics and Institute for Advanced Study, The Hong Kong University of Science and Technology, Hong Kong, China.
[2]Department of Physics and Institute of Computational and Theoretical Studies, Hong Kong Baptist University, Hong Kong, China.

[*] Correspondence to C. T. Chan (phchan@ust.hk)

**ABSTRACT**

It is well known that incident photons carrying momentum $\hbar \mathbf{k}$ exert a positive photon pressure. But if light is impinging from a negative refractive medium in which $\hbar \mathbf{k}$ is directed towards the source of radiation, should light insert a photon "tension" instead of a photon pressure? Using an *ab initio* method that takes the underlying microstructure of a material into account, we find that when an electromagnetic wave propagates from one material into another, the electromagnetic stress at the boundary is in fact indeterminate if only the macroscopic parameters are specified. Light can either pull or push the boundary, depending not only on the macroscopic parameters but also on the microscopic lattice structure of the polarizable units that constitute the medium. Within the context of effective medium approach, the lattice effect is attributed to electrostriction and magnetostriction which can be accounted for by the Helmholtz stress tensor if we employ the macroscopic fields to calculate the boundary optical stress.

**INTRODUCTION**

Electromagnetic (EM) wave or light induced forces have been extensively studied for decades, generating fruitful applications for manipulating microscopic particles, biological materials and macroscopic mechanical components (*1-12*). It is well known that a plane wave in vacuum always exerts a photon pressure on an object in the direction of momentum flux. With the advent of metamaterials (*13-15*) which can have negative refractive indices, can light pressure becomes light tension when light propagates in such a kind of material (*13*)?

For example, consider the situation shown Fig. 1A which depicts a boundary separating two materials, each specified by their own macroscopic permittivity ($\varepsilon$) and permeability ($\mu$). How would the optical stress (e.g. pressure or tension) at the boundary depend on the $\varepsilon$ and $\mu$ of the media on either side of the boundary? The answer to such a question is much more difficult to find than it might seem. Even if assume that the media are truly isotropic and homogenous, we run into difficulties such as the correct interpretation of electromagnetic momentum density (the Abraham/Minkowski controversy) and the related issue of what type of EM stress tensors that should be used in calculations (*16-25*). Another big problem is that metamaterials are always inhomogenous. All metamaterials with unusual optical properties have an underlying microstructure, usually in the form of arrays of subwavelength resonators. In this paper, we will adopt an "*ab initio*" approach and we calculate the boundary stress with a medium treated as an array of coupled scatterers embedded in vacuum with a lattice constant much less than the wavelength. This approach kills two birds with one stone. Firstly, it takes the effect of the microstructure fully into account. Second, it has the advantage that the time-averaged force for a time harmonic field acting on each scatterer (and hence the force density everywhere) can be determined unambiguously using the Maxwell tensor approach as each scattering unit is embedded in air. The boundary stress can hence be determined. We will see the boundary optical stress is in fact indeterminate if we only specify the macroscopic parameters as in Fig. 1A. Light can push or pull on the boundary, depending on the details of the microstructure, although the total force is always positive. We will see that if we want to consider the optical stress problem within the context of effective medium approach, we can use the Helmholtz stress tensor which carries the information of microstructure in the electrostriction and magnetostriction terms. These terms do not appear in standard effective medium theories, but can be calculated with some additional effort (*26*).

**RESULTS**

**Electromagnetic stress at the boundary of the microscopic model system**

Let us start with the simplest possible configuration as shown in Fig.1A, where a boundary separates two semi-infinite materials of different optical properties as specified by their respective permittivity and permeability. An EM plane wave (with electric field amplitude of 1 V/m) propagates from the left-hand side (LHS) to the right-hand side (RHS) and induces a local optical stress on the boundary and we would like to know whether the light exerts a pressure or a tension if the media have unusual [say negative refractive index (NRI)] optical properties. As metamaterials with strange values of refractive indices derive their optical properties from an underlying lattice of scatterers or resonators, a more realistic representation of the boundary region should be the schematic picture shown in Fig. 1B in which the discreteness of the microscopic sub-lattice is explicitly shown. In our calculations, we shall assume that the lattice consist of subwavelength polarizable units schematically shown in Fig.1B, where points of different colors represent units of different materials. For simplicity, we consider a 2D configuration and each polarizable unit is characterized by electric and magnetic polarizabilties $\alpha^e, \alpha^m$. In the case of E-z polarization (electric field along z direction) each unit then represents an out-plane electric monopole $\mathbf{p} = \alpha^e \mathbf{E}$ as well as an in-plane magnetic dipole $\mathbf{m} = \alpha^m \mathbf{H}$. The polarizable units form a lattice (e.g. square) with the size of each unit cell being $a$. In the limit of $\lambda \gg a$, the microscopic lattice can be treated as an effective medium with relative permittivity $\varepsilon_{\text{eff}} = 1 + \alpha^e / (A\varepsilon_0)$ and relative permeability $\mu_{\text{eff}} = (2A + \alpha^m)/(2A - \alpha^m)$, where $A$ denotes the area occupied by a single unit cell (*27*). The EM field within the microscopic lattice system can be determined using multiple scattering theory detailed in Materials and Methods. The force exerted on each polarizable unit can be unambiguously calculated by integrating the time-averaged Maxwell stress tensor $\langle T_{ij}^M \rangle = 1/2 \left[ \varepsilon_0 E_i E_j^* + \mu_0 H_i H_j^* - 1/2 \left( \varepsilon_0 E^2 + \mu_0 H^2 \right) \delta_{ij} \right]$ (*28*) over a closed loop as shown in Fig.1B. We note that the loop resides in vacuum, and hence the Maxwell stress tensor approach is "exact".

In the usual sense of effective medium or homogenization theories as accepted by the optics and metamaterial communities, the microscopic model system depicted in Fig. 1B is macroscopically equal to Fig.1A as long as $\lambda \gg a$. However, some subtleties still arise even in that ideal limit. The boundary in Fig.1A is a geometric line (the black line) but when the microstructure is taken into consideration as in Fig. 1B, where exactly should we draw the boundary? The homogenous system depicted in Fig. 1A is translational invariant along the direction of the boundary, while the discrete system is not and the calculated result will depend on the relative alignment of the left and the right lattice structure. We will use $d$ to denote the relative shift distance between the two sub-lattices with $d = 0$ corresponding to perfect alignment. It is clear in Fig.1B that the boundary encloses a finite region of polarizable units from both sides. To define the boundary stress, this region has to be determined and it can be done in the following way. We will focus on discrete systems with impedance-matched ($\mu_{eff}/\varepsilon_{eff}$ the same for the left and right sub-lattices) configurations and leave the impedance-mismatched cases to later discussion. We first consider an example and choose the values of $\alpha^e$, $\alpha^m$ so that the effectively permittivity and permeability are respectively $\varepsilon_{eff,1} = \mu_{eff,1} = -\varepsilon_{eff,2} = -\mu_{eff,2} = -2$ and the wavelength is fixed to be $\lambda = 100a$ unless otherwise specified. The results obtained with the multiple scattering theory are shown in Fig.1D, where each open circle denotes the calculated force acting on the corresponding polarizable units, with $x = 0$ marking the middle line separating the two lattices. Four configurations with different relative lattice shift $d$ are calculated. We see that for all the configurations, the forces are vanishing small except near the interface region marked by blue color. This is because for impedance-matched configurations, momentum transfer happens at the interface region and it is then natural for us to define the domain within the blue region as the "boundary" and the boundary stress is then defined as the total force per unit length acting on this region. We note that although the force distribution within the interface region does vary with $d$, the total force in the blue domain (i.e., boundary stress) add up to be the same constant as shown in the inserted panel in Fig.1D. This means that the "boundary stress" is well-defined to the extent that it does localize near the geometric boundary and it does not depend on how the micro-lattices are aligned relative to each other.

**Indeterminate stress**

We note that the relationship between a macroscopic effective medium (Fig. 1A) and the underlying microstructure (Fig. 1B) is not a one-to-one mapping. For example, the underlying lattice does not need to be a square lattice as shown in Fig. 1B. We can find a triangular lattice that gives exactly the same effective $\varepsilon_{eff}$ and $\mu_{eff}$. Here, we calculate the boundary stress for two different types of microscopic lattice structures ($C_{4v}$ and $C_{6v}$) that correspond to the same effective macroscopic system in Fig.1A. The results are summarized in Fig.2A-D in which the dots show the calculated boundary stress for different combinations of lattice types as illustrated in the insets. Figure 2A shows the results for the boundary formed by two square sub-lattices, where we fix $n_{\text{eff},1}$ ($n_{\text{eff},1} = \varepsilon_{\text{eff},1} = \mu_{\text{eff},1}$) and vary $n_{\text{eff},2}$ ($n_{\text{eff},2} = \varepsilon_{\text{eff},2} = \mu_{\text{eff},2}$) to see how the stress changes. We see that when $n_{\text{eff},1} = 1$ (corresponding to air), the boundary stress is always negative, meaning that a plane wave propagating in air always pulls the surface of a medium whose microscopic polarizable units form a square lattice. However, for $n_{\text{eff},1} = -1$ the boundary stress can be either positive or negative. Figure 2B shows that the results are very different when the two sub-lattices are triangular. In contrast to the square-lattice case, a plane wave in air ($n_{\text{eff},1} = 1$) always push the boundary. More importantly, the $\sigma_x$ vs. $n_{\text{eff},2}$ relationship is completely different for the two kinds of lattice, even though they are purposely designed to have exactly the same effective $\varepsilon_{\text{eff}}$ and $\mu_{\text{eff}}$. We can also combine a square sub-lattice with a triangular lattice and the results are shown in Fig. 2C with the square lattice on the left and the triangular lattice on the right and Fig. 2D shows the results when the left/right positions are exchanged. We see that the boundary stress changes sign when the positions of the two sub-lattices are exchanged. In addition, we see that the hybrid systems give different boundary stress compared with the previous two cases shown in Fig.2A and B. The four cases considered in Fig.2A-D have very different boundary stresses even though they have by design the same effective macroscopic permittivity and permeability. This can be illustrated in Fig.2E where the case with $n_{\text{eff},1} = -1, n_{\text{eff},2} = 4$ is taken as an example. The four types of lattice correspond to the same macroscopic effective system as depicted in the center panel. And yet, the stress exerted on the boundary is different in sign (indicated by the arrow) and magnitude in the four cases. These results show that the light induced boundary stress for a metamaterial system depicted in Fig. 1A is indeterminate even if the macroscopic permittivities and permeabilities are completely specified. The results depend on microscopic details. We will see that this rather counter-intuitive phenomenon is due to the existence of electrostricton and magnetostriction effects in the second part of the paper.

If the sub-lattices' impedances do not match each other, the wave reflection at the boundary will result in a field gradient that induces a gradient force on each polarizable unit. In that case, the boundary stress is not necessarily well-defined as the optical force is not localized near the boundary and there is no unique way to attribute part of the force to the boundary. However, if the LHS sub-lattice is air (see the insets of Fig.3), the boundary stress is still well-defined even if the surface impedance of the RHS half space is not matched with air. In this case, the reflected field on the LHS does not affect the boundary stress. Here we consider $n_{eff,2} < 0$ with $\varepsilon_{eff,2} \neq \mu_{eff,2}$ so that its impedance does not match that of air on the LHS. The dots in Fig. 3A and B show respectively the boundary stress for the square lattice and triangle lattice, where we fix $\varepsilon_{eff,2} = -1$. It is seen that the stress is pulling in the square-lattice but is pushing in the triangle lattice case. If we fix $\mu_{eff,2} = -1$ and change $\varepsilon_{eff,2}$ the results are similar (Fig.3C,D). Hence, even in the general case of an impedance-mismatched boundary, whether the light exerts a pressure or a tension on the boundary depends on microscopic details even though the macroscopic effective medium parameters are the same.

We note that if the microscopic system is a slab of finite thickness, the total force induced on the system by an incident plane wave is the same for different lattice types, as long as their effective permittivity and permeability have the same value. Take for example the configuration depicted in Fig.1B, and suppose that the medium on both sides of the lattice (yellow and green colored regions) is air. This system corresponds to a finite-sized slab. Again we calculate the force acting on each polarizable unit in this case. The dots in Fig. 4A and B show the internal force distributions for a square-lattice and a triangle-lattice systems, respectively. We have set $\varepsilon_{\text{eff},1} = \mu_{\text{eff},1} = -1$ and $\varepsilon_{\text{eff},2} = \mu_{\text{eff},2} = 4$. We see that only the polarizable units at the boundary feel a force due to impedance-matched configuration and the boundary force is very different for two kinds of lattice. In Fig. 4C, we show the cumulative force summed starting from the left-most layer of the slab and the final value (ending on the RHS and marked by a blue dot) gives the total force acting on the slab. The total forces (blue dot) are equal for both cases of Fig.4A and B while the forces on individual boundaries are different. Figure 4C and D show an impedance-mismatched case with $\varepsilon_{\text{eff},1} = \mu_{\text{eff},1} = -1$ and $\varepsilon_{\text{eff},2} = 1, \mu_{\text{eff},2} = 4$, where the optical forces extend into the bulk due to the field gradient arising from Fabry-Perot reflections. We see that the force distributions in the bulk and at the interface regions are very different for different lattices. In this case, the boundary stress is hard to define but we see from Fig. 4F that the total force (end value of the lines in Fig. 4F marked by the blue dot) again adds up to the same value for the two lattice configurations. These results show that the total forces acting on an object is the same as long as the macroscopic parameters are the same, but the internal force can be very different in sign and magnitude for different underlying lattices. We further consider the total force exerting on a semi-infinite reflecting metallic material when the plane wave is incident from a NRI material. The metallic material on the right of the boundary is modelled by a microscope lattice structure which gives a negative effective permittivity and $\mu_{\text{eff}} = 1$ in the effective medium limit so that the field decays exponentially into the bulk of the metallic material. The results for a square lattice and a triangle lattice are shown in Fig.4G and H, respectively. We have set $\varepsilon_{\text{eff},1} = \mu_{\text{eff},1} = \pm 1$ (i.e. air and NRI) and $\varepsilon_{\text{eff},2} = -4$, $\mu_{\text{eff},2} = 1$. It is noted that no matter the reflecting metal is forming an boundary with air or NRI, the total forces (sum of the forces acting on all points with $x > 0$) exerting on the metallic material are always positive for both the square lattice and triangle lattice. These results show that light always pushes on a metallic material, independent of whether it is incident from a positive or negative index medium.

We now consider the situation in which the underlying lattice is amorphous as shown in Fig.5A where both sub-lattices consist of randomly positioned polarizable units. In this case, we employ a numerical solver COMSOL (*29*) for calculating the fields and hence the boundary stress. In the simulations, each polarizable unit is defined by a cylinder with radius $r$, relative permittivity $\varepsilon_c$ and relative permeability $\mu_c$. In the long wavelength limit and E-z polarization, the system is essentially a collection of in-plane magnetic dipoles and out-of-plane electric monopoles with effective parameters given by $\varepsilon_{\text{eff}} = 1 + \rho(\varepsilon_c - 1), \mu_{\text{eff}} = -1 + 2/(1 - \rho M)$, where $M = (\mu_c - 1)/(\mu_c + 1)$ and $\rho$ is the filling ratio. We limit our discussions in the impedance-match cases and other cases can be considered similarly. Due to the fluctuations induced by the randomness, the force acting on each cylinder is generally nonzero and shows fluctuation even for those located outside the boundary region, but the boundary stress can still be meaningfully determined if we average out the fluctuation. We calculate the stress by averaging the total force per unit length over domains of different sizes (each containing the same number of cylinders from the two sub-lattices) centered at the geometric interface (see Fig.S2 in Supplementary Materials). This process is carried out for 10 sample realizations for each combination of $n_{\text{eff},1}$ and $n_{\text{eff},2}$, and the boundary stress is taken to be an ensemble average. For each sample configuration, each sub-lattice consists of 300 cylinders randomly distributed in an area of $l \times w = 30a \times 10a$. The radii of the two kinds of cylinders are set to be $r_1 = 0.1a$ and $r_2 = 0.16a$.

Periodic boundary condition is applied on the upper and lower boundaries of the "supercell". We fix $n_{\text{eff},1} = 2$ and vary $n_{\text{eff},2}$. The yellow circles in Fig.5B denote the boundary stress calculated by COMSOL for the 10 realizations and the dotted yellow line is the ensemble average. Figure 5B also shows for comparison the boundary stress of the square lattice, denoted as blue circles. A clear difference between the amorphous lattice and the square lattice is noted. To take a further step, we studied the transition of the boundary stress from a square lattice to an amorphous lattice. Let $\delta$ be the random amplitude which characterizes the deviation of a cylinder from the unit-cell center of the square lattice ($\delta = 0$ corresponds to the unperturbed square lattice). For all the cases the distance between two arbitrary cylinders meets the condition $d \geq 0.8a$ so that the cylinders will not come too close together. The results are shown in Fig.5B, each circle denotes the boundary stress of a sample and dotted lines are the ensemble averages, which shows that the boundary stress gradually approaches that of the amorphous configuration as $\delta$ is increased. It is clear from the above results that the amorphous-lattice system gives a different boundary stress compared to that of a square-lattice system.

**Electromagnetic stress at the boundary from an effective-medium perspective**

The results in Figs. 2-5 show that specifying the standard effective-medium parameters ($\varepsilon_{\text{eff}}$, $\mu_{\text{eff}}$) does not provide sufficient information to determine the boundary optical stress. We iterate that there is nothing "wrong" *per se* in the effective medium description. The lattice system in Fig.1B, with $\lambda = 100a$, can be well represented as an effective-medium system (Fig.1A). This can be justified by the comparison of the $E_z$ field in the two kinds of system shown in Fig.1C, where we considered square lattice with $\varepsilon_{\text{eff},1} = \mu_{\text{eff},1} = -\varepsilon_{\text{eff},2} = -\mu_{\text{eff},2} = -2$ (the dots in Fig.1C mark the position of the polarizable units). The field shows very similar phase and magnitude distribution in the two systems. It can be further improved if we go to a smaller a/λ ratio, but the lattice structure information is still required to describe the boundary stress. If we want to stick with an effective medium approach, and avoid the burden of doing a full wave *ab initio* calculation taking the lattice structure explicitly into account (either multiple scattering or with numerical solvers), we need a EM stress tensor formulation that takes into account the lattice symmetry. Such a formulation exists. The Helmholtz stress tensor (*30-34*) incorporates the underlying symmetry information through the electrostriction and magnetostriction effects. For a lattice system it takes the expression (*26, 30*):

$$\left\langle T_{ij}^{H} \right\rangle = \frac{1}{2} \text{Re} \begin{bmatrix} \varepsilon_0 \varepsilon_{\text{eff}} E_i E_j^* + \mu_0 \mu_{\text{eff}} H_i H_j^* - \frac{1}{2}\left(\varepsilon_0 \varepsilon_{\text{eff}} E^2 + \mu_0 \mu_{\text{eff}} H^2\right)\delta_{ij} \\ -\frac{1}{2}\left(\varepsilon_0 \frac{\partial \varepsilon_{\text{eff}}}{\partial u_{ij}} E^2 + \mu_0 \frac{\partial \mu_{\text{eff}}}{\partial u_{ij}} H^2\right) \end{bmatrix}, \quad (1)$$

where $u_{ij}$ is the strain tensor and $E$, $H$ are the fields in the corresponding effective-medium system. The electrostriction and magnetostriction terms ($\partial \varepsilon_{\text{eff}} / \partial u_{ij}, \partial \mu_{\text{eff}} / \partial u_{ij}$) are explicitly lattice-symmetry dependent. They can be analytically derived by multiple scattering theory or numerically retrieved from the band diagram of the lattice (*26*). For 2D system and E-z polarization one can show that:

$$\frac{\partial \varepsilon_{\text{eff}}}{\partial u_{xx}} = \frac{\partial \varepsilon_{\text{eff}}}{\partial u_{yy}} = 1 - \varepsilon_{\text{eff}}, \quad \frac{\partial \mu_{\text{eff}}}{\partial u_{xx}} = -\frac{\mu_{\text{eff}}^2 - 1}{2} + \gamma \frac{(\mu_{\text{eff}} - 1)^2}{2} \cos 2\phi,$$

$$\frac{\partial \mu_{\text{eff}}}{\partial u_{yy}} = -\frac{\mu_{\text{eff}}^2 - 1}{2} - \gamma \frac{(\mu_{\text{eff}} - 1)^2}{2} \cos 2\phi, \quad \frac{\partial \mu_{\text{eff}}}{\partial u_{xy}} = \frac{\partial \mu_{\text{eff}}}{\partial u_{yx}} = \nu \frac{(\mu_{\text{eff}} - 1)^2}{2} \sin 2\phi, \quad (2)$$

where $\phi$ is the angle between the direction of the local wave vector and *x* axis (for the normal incidence $\phi = 0$). The coefficients $\gamma$ and $\nu$ are symmetry-dependent parameters and for a square lattice of *C*4v symmetry $\gamma_{\text{square}} = 1.297, \nu_{\text{square}} = -0.596$ while for a triangular lattice of *C*6v symmetry $\gamma_{triangle} = 0.5, \nu_{triangle} = 0$ (*26*). With this stress tensor we can analytically determine the boundary stress as

$$\sigma_x = \frac{1-\gamma}{8}\left[\left(\mu_{eff,2}-1\right)^2 - \left(\mu_{eff,1}-1\right)^2\right]\mu_0 H^2 \tag{3}$$

with $\gamma = \gamma_{square}$ for the square-lattice system and $\gamma = \gamma_{triangle}$ for the triangle-lattice system. Here $H$ is the magnetic field at the boundary. For hybrid system formed by a square sub-lattice on the left and a triangle sub-lattice on the right, the boundary stress is expressed as

$$\sigma_x = \frac{1}{8}\left[\left(1-\gamma_{triangle}\right)\left(\mu_{eff,2}-1\right)^2 - \left(1-\gamma_{square}\right)\left(\mu_{eff,1}-1\right)^2\right]\mu_0 H^2 . \tag{4}$$

For the triangle-square case the positions of $\gamma_{square}$ and $\gamma_{triangle}$ are interchanged in the above equation. Equations (3) and (4) show that on one hand that the boundary stress can be determined by the macroscopic field obtained using standard effective parameters but the coefficient $\gamma$ arising from the underlying lattice structure must be known in order for the boundary stress to be determinate. The solid lines in Figs.2 and 3 mark the boundary stress calculated analytically with Eqs. (3) and (4), which agree very well with that of the microscopic model system (dots).

The Helmholtz stress tensor for the amorphous system is (*30*):

$$\langle T_{ij} \rangle = \frac{1}{2}\text{Re}\left[\begin{array}{c} \varepsilon_0\varepsilon_{eff} E_i E_j^* + \mu_0\mu_{eff} H_i H_j^* - \frac{1}{2}\left(\varepsilon_0\varepsilon_{eff} E^2 + \mu_0\mu_{eff} H^2\right)\delta_{ij} \\ +\frac{1}{2}\left(\varepsilon_0\rho\frac{\partial\varepsilon_{eff}}{\partial\rho}E^2 + \mu_0\rho\frac{\partial\mu_{eff}}{\partial\rho}H^2\right)\delta_{ij} \end{array}\right], \tag{5}$$

where the electrostriction and magnetostriction terms depend on the filling ratio $\rho$ and : $\partial\varepsilon_{eff}/\partial\rho = (\varepsilon_{eff}-1)/\rho, \partial\mu_{eff}/\partial\rho = (\mu_{eff}^2-1)/(2\rho)$. Appling the above stress tensor one can analytically obtain the boundary stress with the same expression as Eq.(3) (with $\gamma = 0$). The result is shown in Fig.5B (red lines) and a good agreement with the result of the microscopic amorphous system is again noted.

We showed in the above discussions that it is possible to reproduce the boundary stress of the microscopic system within the context of effective medium theory. To do this one needs to know the microscopic detail of the lattice and derive the corresponding electrostriction and magnetostriction corrections incorporated in the Helmholtz stress tensor for evaluation of the boundary stress. Expressions of Eqs. (2-5) for the H-z polarization can also be obtained by invoking the duality relationship and substitute $\mu \to \varepsilon$ and $H \to -E$ and an example is given in Fig.S1 in the Supplementary Materials. We note that while the Helmholtz stress tensor can predict the total boundary stress, it obviously cannot reproduce the finer details at the boundary such as the dependence of the local force density distribution on the relative shift of the lattice on the left and right as shown in Fig. 1D. There is simply no information within the context of effective medium that would allow the determination of such details.

**Electromagnetic stress at the boundary of a metamaterial system**

We have so far considered idealized metamaterials in which the underlying microstructures are point dipoles arranged in a lattice with $a \ll \lambda$. We do this on purpose so that the results we obtained cannot be attributed to the "coarse-grainedness" of the lattice or the complexity of the internal structure of the resonators, but rather to something more intrinsic. In experimentally realizable metamaterials, the internal dimension of the resonators is usually not small compared with $a$, and $a$ is not that small compared with wavelength. For completeness, we also consider such configurations here by considering a model 2D metamaterial system composing a periodic array of core-shell cylinders (Fig.6A) with a perfect electric conductor core of radius $r_{inner} = 0.144a$ and a high dielectric outer shell of $r_{outer} = 0.272a$ and $\varepsilon_{shell} = 50, \mu_{shell} = 1$. Such a system can be used to realize an isotropic NRI material under the H-z polarization (*35, 36*). The band diagram for a square lattice of such core-shells is shown in Fig.6B, which shows that the second band has a negative group velocity. Using standard retrieval procedures (*37, 38*), we obtained the effective parameters corresponding to this band which is shown in Fig.6C. The $\varepsilon_{\text{eff}}$ and $\mu_{\text{eff}}$ are both negative so that the system behaves as an NRI material at that frequency range. For force calculations, we use 10 layers of core-shell cylinders on the LHS to represent the NRI material and equal area of positive-refractive-index material on the RHS represented by a fine-grained array (lattice constant = $a/10$) of dielectric cylinders of radius $r_c = 0.02a$, as shown in Fig. 6A. Periodic condition in the *y* direction is assumed. We pick the normalized frequency $\omega = 0.327$ at which $\varepsilon_{\text{eff}} = \mu_{\text{eff}} = -0.135$ for the core-shell sub-lattice and calculate the boundary stress while varying $\varepsilon_c, \mu_c$ of the cylinders on the RHS. We note that at this frequency and with $r_{\text{outer}} = 0.272a$, we have a "coarse-grained" configuration and the size of core-shell cylinders is not small compared with *a*. The results of the boundary stress are shown in Fig. 6D, which shows that it is positive for square lattice. We then repeat the calculation for a triangular-lattice arrangement (for both LHS/RHS), with the lattice constant adjusted so that $\varepsilon_{\text{eff}} = \mu_{\text{eff}} = -0.135$ at the same frequency. As shown in Fig. 6D, the stress now becomes negative. The dots in the figure represent the numerical results of the microscopic systems while the red lines denote the analytical results predicted by using the H-z version of Eq.(3). We note a good agreement between the two. Figure 6E shows the $H_z$ field distribution for the square-lattice case with $n_{\text{eff},2} = 1$ (air), where based on the phase variation we confirm that the Bloch wave vector does satisfy the condition of $k_{\text{bloch}} a \ll 1$, which is a necessary condition for the effective medium theory to work.

**DISCUSSION**

In summary, we studied the EM stress induced at the boundary formed by two kinds of materials using a generic microscopic model. We gave explicit examples to show that square, triangle and amorphous lattices which are designed to have the same macroscopic $\varepsilon_{eff}$ and $\mu_{eff}$ give very different boundary optical stresses, both in sign and in magnitude. One hence cannot tell whether the optical stress at the boundary region is compressive or tensile even if macroscopic $\varepsilon_{eff}$ and $\mu_{eff}$ are completely specified. The underlying lattice structure strongly affects the boundary stress and one can determine whether the boundary is pushed or pulled only if the microscopic lattice symmetry is specified in addition to $\varepsilon_{eff}$ and $\mu_{eff}$.

We need to emphasize again here that the indeterminate value of the boundary stress is not due to the inaccuracy of the effective permittivity and permeability. In the numerical results we presented, we considered $\lambda=100a$ and in this small $a/\lambda$ regime, $\varepsilon_{eff}$ and $\mu_{eff}$ are very accurate in the usual sense of effective medium or homogenization theories. The lattice structure dependence will persist even if $a/\lambda$ tends to an arbitrarily small number. Such a dependence can be understood from the perspective of the virtual work principle. As the optical stress can in principle be determined by calculating the change of free energy under a virtual deformation of the lattice (*30*), the strain tensor and its underlying symmetry should come into the expression of the stress tensor, which manifests in the form of electrostriction and/or magnetostriction correction within the context of an effective medium description. When treating such a system with effective medium theory, the Helmholtz stress tensor can be used to predict correctly the boundary stress as the symmetry information is taken into account as electrostriction and magnetostriction terms in the Helmholtz tensor formulation. Other forms of stress tenors such as Abraham, Minkowski or Einstein-Laub cannot predict the correct results for boundary optical stresses due to the absence of lattice-structure dependent terms (see Materials and Methods). We note that our conclusion applies to all metamaterials that have been made in practice, as all such materials are artificial composites composing of an underlying structure.

Last but not least, we note that the macroscopic effective parameters do provide sufficient information to determine the total optical force acting on a finite-sized object and the total force does not depend on the details of the microstructure as long as the effective parameters are accurate which is usually the case when a/λ is small. It is the stress on the boundary that is indeterminate if only the macroscopic parameters are specified. If an object is immersed in a NRI medium, the total optical force is always positive (i.e. light pushes the object) while the boundary stress can be either a pressure or a tension depending on the details of the microstructure that makes up the NRI medium. In some special configurations, a beam of light can attract an object (*6, 39-44*) but that is due to the special properties of the light beam rather than the refractive index of the medium in which the beam travels. In addition to metamaterials, the current study may also find applications in the field of biophysics (e.g., how the EM stress will deform a biological cell). The relationship between microscopic lattice symmetry and boundary stress in three dimensions also warrants further study in the future.

**MATERIALS AND METHODS**

**Multiple scattering formulation of the microscopic model**

The multiple scattering theory is formulated here for the E-z polarization and the H-z polarization can be obtained easily by invoking duality relations. We first assume that the semi-infinite homogenous medium on both sides of the lattice in Fig.1B are air, in which case the system corresponds to a slab of metamaterial. Denote the field acting on the polarizable unit locating at $i$th column and $j$th row as $\mathbf{E}_{ij}$ and $\mathbf{H}_{ij}, i = 1, 2, ..., N_1 + N_2$ with $N_1, N_2$ being the number of column layers of the two sub-lattices. The induced dipole/monopole moments are $\mathbf{p}_{ij} = \alpha_i^e \mathbf{E}_{ij}$, $\mathbf{m}_{ij} = \alpha_i^m \mathbf{H}_{ij}$ and the following self-consistent equation can be formulated

$$\sum_{(i',j')} \begin{bmatrix} \delta_{ii'}\delta_{jj'}/\alpha_i^e \bar{\mathbf{I}} - \mu_0\omega^2 \bar{\mathbf{G}}(\mathbf{r}_{i'j'}, \mathbf{r}_{ij}) & -i\omega\nabla\times\bar{\mathbf{G}}(\mathbf{r}_{i'j'}, \mathbf{r}_{ij}) \\ i\omega\nabla\times\bar{\mathbf{G}}^m(\mathbf{r}_{i'j'}, \mathbf{r}_{ij}) & \delta_{ii'}\delta_{jj'}/\alpha_i^m \bar{\mathbf{I}} - \varepsilon_0\omega^2 \bar{\mathbf{G}}^e(\mathbf{r}_{i'j'}, \mathbf{r}_{ij}) \end{bmatrix} \begin{bmatrix} \mathbf{p}_{i'j'} \\ \mathbf{m}_{i'j'} \end{bmatrix} = \begin{bmatrix} \mathbf{E}_{ij}^{\text{inc}} \\ \mathbf{H}_{ij}^{\text{inc}} \end{bmatrix} \quad (6)$$

where $\delta_{ii'}$ denotes the Kronecker delta; $\mathbf{r}_{ij}, \mathbf{r}'_{ij}$ are the position vectors of the target and source units, respectively. $\bar{\mathbf{G}}(\mathbf{r}_{ij}, \mathbf{r}_{i'j'}) = (\bar{\mathbf{I}} + \nabla\nabla/k^2)g(\mathbf{r}_{ij} - \mathbf{r}_{i'j'})$ is the dyadic Green's function with $g(\mathbf{r}_{ij}, \mathbf{r}_{i'j'}) = i/4 H_0^{(1)}(k|\mathbf{r}_{ij} - \mathbf{r}_{i'j'}|)$ and $H_0^{(1)}(z)$ being the zeroth order cylindrical Hankel function of the first kind. Substitute the expression of Green's function into the above equation and employ the periodic condition in the $y$ direction, we obtain

$$\sum_{i'=1}^{N_1+N_2} \begin{bmatrix} \delta_{i,i'} - \mu_0\omega^2\alpha_{i'}^e F_1(n,d) & i\omega\mu_0\alpha_{i'}^m F_3(n,d) & -i\omega\mu_0\alpha_{i'}^m F_2(n,d) \\ i\omega\alpha_{i'}^e F_3(n,d) & \delta_{i,i'} + \alpha_{i'}^m F_6(n,d) & -\alpha_{i'}^m F_5(n,d) \\ -i\omega\alpha_{i'}^e F_2(n,d) & -\alpha_{i'}^m F_5(n,d) & \delta_{i,i'} + \alpha_{i'}^m F_4(n,d) \end{bmatrix} \begin{bmatrix} E_{i'}^z \\ H_{i'}^x \\ H_{i'}^y \end{bmatrix} = \begin{bmatrix} E_i^{\text{inc},z} \\ 0 \\ H_i^{\text{inc},y} \end{bmatrix} \quad (7)$$

where $n = i - i'$ and $d$ denotes the transverse shift of the target layer with respect to the source layer along $y$ direction. $F_i(n,d)$ are lattice sums that take the following expressions (45):

$$F_1(n,d) = \sum_m \frac{i}{2bq_m} e^{i|n|aq_m} e^{i(k_y - K_m)d}, \quad F_2(n,d) = \text{sign}(n)\sum_m \frac{-1}{2b} e^{i|n|aq_m} e^{i(k_y - K_m)d},$$

$$F_3(n,d) = \sum_m \frac{K_m - k_y}{2bq_m} e^{i|n|aq_m} e^{i(k_y - K_m)d}, \quad F_4(n,d) = \sum_m \frac{-iq_m}{2b} e^{i|n|aq_m} e^{i(k_y - K_m)d},$$

$$F_5(n,d) = \text{sign}(n)\sum_m \frac{i(K_m - k_y)}{2b} e^{i|n|aq_m} e^{i(k_y - K_m)d}, \quad F_6(n,d) = \sum_m \frac{-i(K_m - k_y)^2}{2b} e^{i|n|aq_m} e^{i(k_y - K_m)d},$$

(8)

where $K_m = 2\pi m / a$, $K_x = q_0$ and $k^2 = q_m^2 + (k_y - K_m)^2$ with $k_y$ being the Bloch wave vector along $y$ direction. $m$ is the diffraction order and the summation over $m$ is truncated to certain value in implementation. $\text{sign}(n)$ is the sign function. Substitute the above equation into Eq.(7) and one can solve the resulted linear equations for the fields at the positions of the polarizable units.

If homogenous medium instead of air are attached to the lattice on both sides as in Fig.1B, corresponding to the case that the boundary is formed by two semi-infinite materials, both the incident and the scattered fields from the dipoles are multi-reflected by the boundaries of the medium. This becomes a plane wave transmission problem because each channel of the scattered fields of the dipoles/monopoles [i.e., the lattices sums in Eq. (8)] takes a simple plane wave form, and the results can be found in many textbooks such as Ref.(*46*). The final form of the self-constant equations are similar to Eq. (7) except that the lattice sums now have extra contributions from the multiple reflections due to the two boundaries. Note that the boundary stress is not affected by the width of lattice as long as enough number of column layers are considered.

Once the field acting on the polarizable units are obtained, the induced dipole/monopole moments $\mathbf{p}_{ij}, \mathbf{m}_{ij}$ can be calculated straightforwardly and the field at arbitrary position can also be easily obtained employing the Green's function. The optical force acting each polarizable unit is then evaluated by integrating the Maxwell stress tensor over a closed loop as shown in Fig.1B.

**Numerical simulation**

For the amorphous-lattice system, we do not have a semi-analytical method that can be used to calculate the boundary stress. So we performed the full-wave EM simulations with a commercial finite-element package COMSOL Multiphysics (*29*). The positions of the cylinders are first generated by a random number generator and then imported into COMSOL to form the model geometry. Periodic boundary condition is applied in the $y$ direction for the supercell while scattering boundary condition is applied in the *x* direction, with a background plane wave excited from the LHS. The EM force acting on each point is then calculated by the Maxwell stress tensor method as a post-processing step.

**Boundary stress given by Abraham, Minkowski and Einstein-Laub tensors**

In homogeneous and isotropic medium, the time-averaged Abraham and Minkowski tensors take equal form of (*16-18, 24*)

$$\langle T_{ij}^{\text{AM}} \rangle = \frac{1}{2}\text{Re}\left[\varepsilon_0 \varepsilon E_i E_j^* + \mu_0 \mu H_i H_j^* - \frac{1}{2}\left(\varepsilon_0 \varepsilon E^2 + \mu_0 \mu H^2\right)\delta_{ij}\right] \quad (9)$$

while Einstein-Laub tensor writes

$$\langle T_{ij}^{\text{EL}} \rangle = \frac{1}{2}\text{Re}\left[\varepsilon_0 \varepsilon E_i E_j^* + \mu_0 \mu H_i H_j^* - \frac{1}{2}\left(\varepsilon_0 E^2 + \mu_0 H^2\right)\delta_{ij}\right]. \quad (10)$$

For the E-z polarization of the effective-medium system, the former contributes to a boundary stress of

$$\sigma_x = \langle T_{xx}^{\text{AM},2} \rangle - \langle T_{xx}^{\text{AM},1} \rangle = -\frac{1}{4}\left[\left(\varepsilon_{\text{eff},2} - \varepsilon_{\text{eff},1}\right)\varepsilon_0 E^2 + \left(\mu_{\text{eff},2} - \mu_{\text{eff},1}\right)\mu_0 H^2\right], \quad (11)$$

which does not have a dependence on the lattice symmetry and predicts the same result for different microscopic lattice systems. The latter simply vanishes at the boundary due to the continuity of the parallel field components. Hence none of the Abraham tensor, Minkowski tensor or Einstein-Laub tensor can explain the boundary stress phenomena in the main text.

**Acknowledgements:** We thank Prof. Z. Q. Zhang and Dr. W. J. Sun for their valuable comments and suggestions. **Funding:** This work was supported by AOE/P-02/12. **Author contributions:** S.B.W. did the calculations. J.N. helped on the theory of Helmholtz stress tensor. M.X. helped on the Green's function method. C.T.C. conceived the idea and supervised the project. S.B.W. and C.T.C co-wrote the manuscript. **Competing interests:** The authors declare no competing financial interests. **Data and materials availability:** All data presented in this work are available upon request to S.B.W.

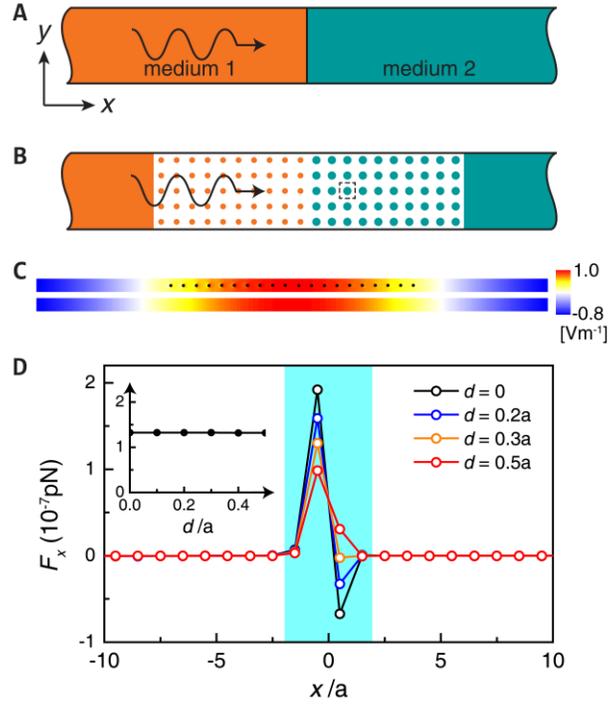

**Fig. 1. Boundary optical stress in the microscopic model system.** (**A**) A boundary formed by two different homogenous media described by effective parameters. (**B**) For metamaterials, the effective medium has an underlying microscopic structure, taken to be a square lattice in this example. (**C**) The effective-medium [panel (**A**)] is a good representation of the microscopic system [panel (**B**)] if the fields calculated for them are nearly the same as shown in this panel where we compare the calculated electric fields ($E_z$). (**D**) Force distribution in the model system with $\varepsilon_{\text{eff},1} = \mu_{\text{eff},1} = -\varepsilon_{\text{eff},2} = -\mu_{\text{eff},2} = -2$ and $\lambda = 100a$ with $a$ being the size of the unit cell. $d$ is the relative shift between two sub-lattices along $y$ direction. The inserted panel shows the total force (boundary stress) acting on the boundary, which is the sum of the non-vanishing force values in the blue shadowed region.

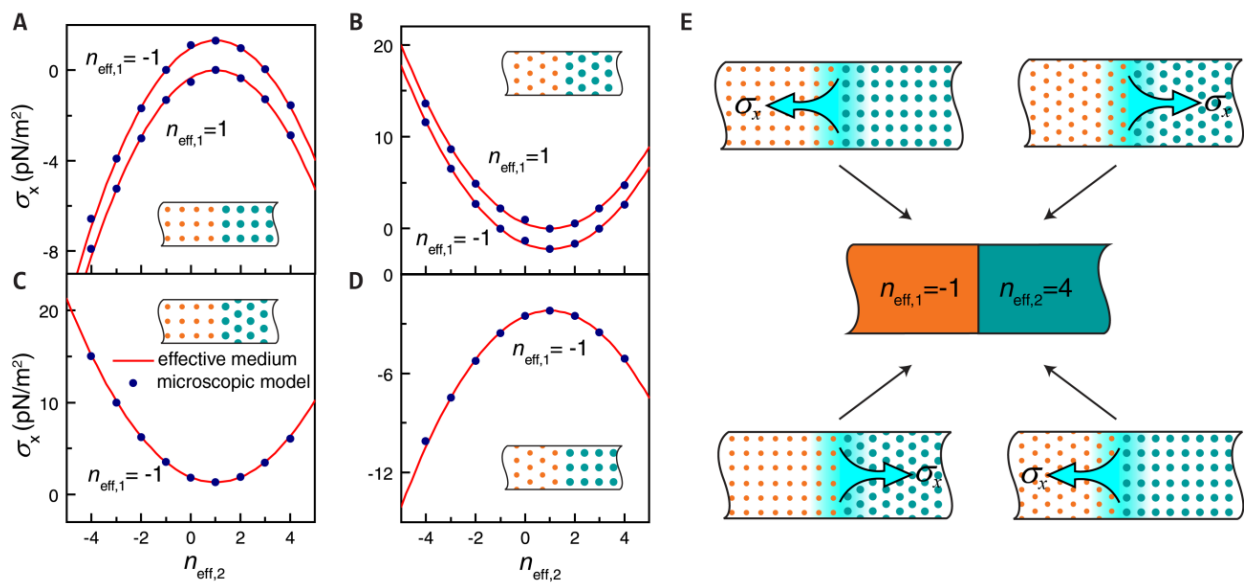

**Fig. 2. Boundary optical stress in impedance-matched lattice systems.** Boundary stress in the square-square (**A**), triangle-triangle (**B**), square-triangle (**C**) and triangle-square (**D**) systems. We fix the refractive index on the left $n_{\text{eff},1} = -1$ or $n_{\text{eff},1} = 1$ (corresponding to air) and vary the refractive index $n_{\text{eff},2}$ on the RHS. The dots correspond to the boundary stresses of the microscopic systems calculated with multiple scattering theory while the lines denote the analytical values obtained using the Helmholtz stress tensor. The insets show the configurations under consideration. (**E**) Four types of lattice systems correspond to the same effective-medium system ($n_{\text{eff},1} = -1$ on the left, $n_{\text{eff},2} = 4$ on the right) but have entirely different boundary stresses. The direction of the stresses is shown in the panels by an arrow.

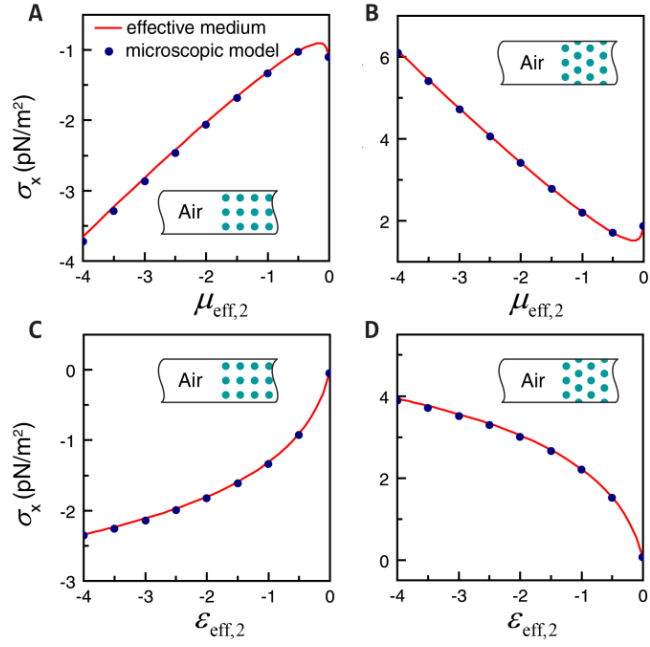

**Fig. 3. Boundary optical stress in impedance-mismatched lattice systems.** (**A**) and (**C**) show the boundary stress as a function of $\mu_{\text{eff},2}$, $\varepsilon_{\text{eff},2}$ in the "air/square-lattice" configuration, respectively. (**B**) and (**D**) show the boundary stress as a function of $\mu_{\text{eff},2}$, $\varepsilon_{\text{eff},2}$ in the "air/triangular-lattice" configuration, respectively. In (**A**) and (**B**) we fix $\varepsilon_{\text{eff},2} = -1$ while in (**C**) and (**D**) we fix $\mu_{\text{eff},2} = -1$.

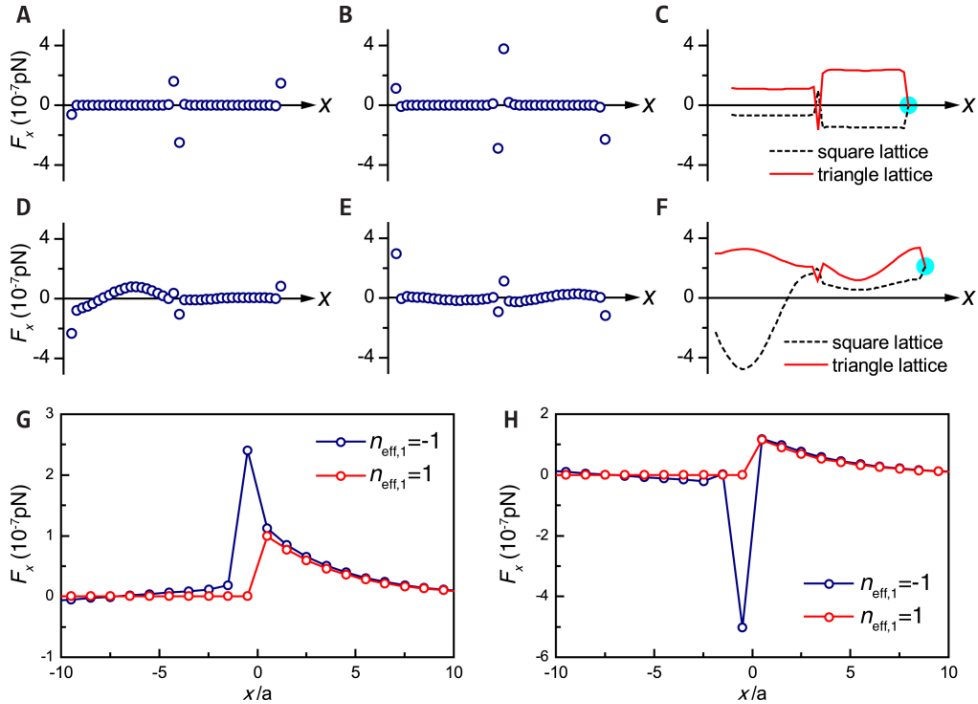

**Fig. 4. Optical force distribution in finite-slab and total-reflecting systems.** Optical force distribution in finite square-lattice [(**A**), (**D**)] and triangle-lattice [(**B**), (**E**)] slab systems. For (**A**) and (**B**) the impedances of the lattice systems match that of air while for (**C**) and (**D**) they does not. The solid and dashed lines in (**C**) and (**F**) denote the accumulated force counting from the first layer and the final value (as marked by a blue dot) corresponds to the total force exerting on the slab. We note that while the forces on the boundary depend strongly on the microscopic details, the total force acting on the slab is exactly the same independent of the microscopic details as long as the macroscopic $\varepsilon_{\text{eff}}$ and $\mu_{\text{eff}}$ are the same. Optical force distribution for a semi-infinite square (**G**) [triangle (**H**)] lattice with $\varepsilon_{\text{eff},2}=-4$, $\mu_{\text{eff},2}=1$ attached to a semi-infinite square (triangle) lattice with $\varepsilon_{\text{eff},1}=\mu_{\text{eff},1}=-1$ or air with $\varepsilon_{\text{eff},1}=\mu_{\text{eff},1}=1$. Here $x$ denotes the position of the unit with $a$ being the column layer distance.

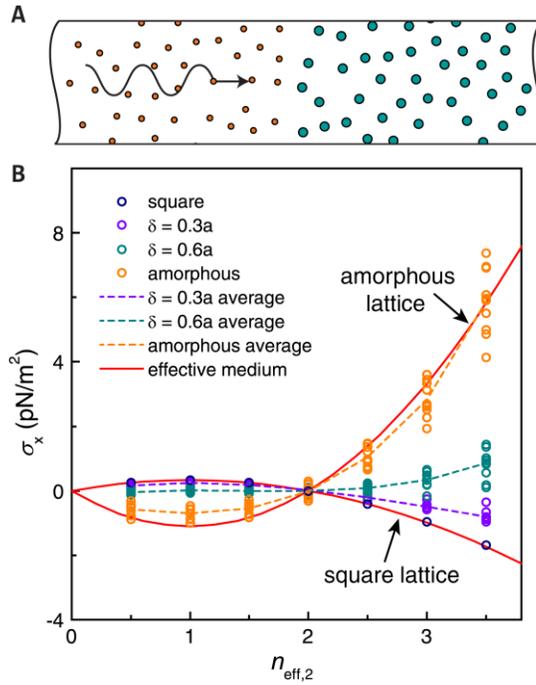

**Fig. 5. Boundary optical stress in amorphous-lattice system.** (**A**) A part of an amorphous-lattice sample is drawn to scale. (**B**) Comparison between the boundary stress of a square lattice and an amorphous lattice. Circles denote the stress calculated numerically for the microscopic lattice systems. Solid lines denote the stress calculated analytically using the Helmholtz stress tensor in the effective medium system. Dashed lines are the average of the values represented by circles. $\delta$ is the random amplitude defined in the main text.

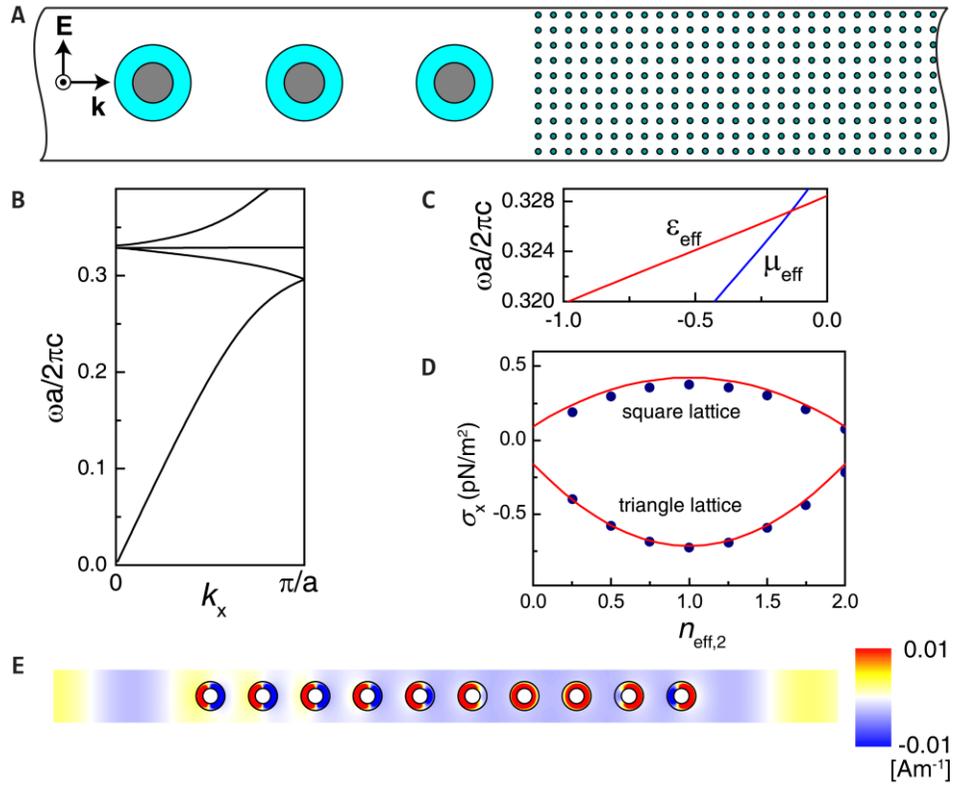

**Fig. 6. Boundary optical stress in a metamaterial system.** (**A**) Metamaterial system formed by core-shell cylinders and normal dielectric cylinders. (**B**) Photonic band diagram for the core-shell square lattice. (**C**) Retrieved effective permittivity and permeability based on S-parameters of the core-shell structure. (**D**) Boundary stresses of the microscopic model system (dots) and the corresponding effective-medium system (lines). (**E**) Magnetic field distribution in the $n_{\text{eff},2}=1$ (air) case.

Supplementary Materials for

**Electromagnetic stress at the boundary: photon pressure or tension?**

Shubo Wang, Jack Ng, Meng Xiao, and C. T. Chan

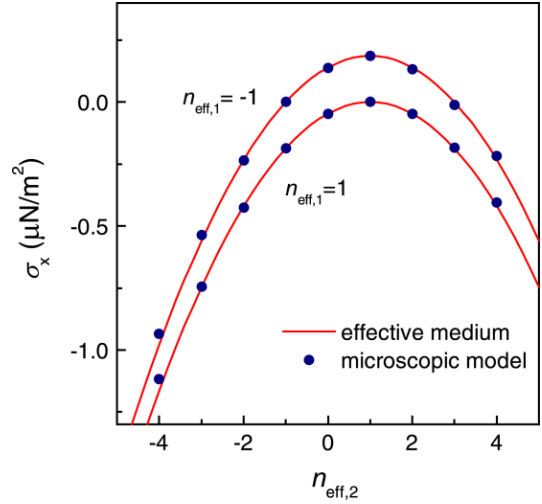

**Fig. S1. Boundary stress in square-lattice system under H-z polarization.** Dots represent the results of the microscopic model system while lines denote the results given by Helmholtz stress tensor in the effective medium system. We consider that the two sub-lattices with $n_{\text{eff},1} = \varepsilon_{\text{eff},1} = \mu_{\text{eff},1}$ and $n_{\text{eff},2} = \varepsilon_{\text{eff},2} = \mu_{\text{eff},2}$. The incident wave has a magnetic field amplitude of 1 A/m. The result of the effective-medium system is obtained through Eq.(3) in the main text and invoking duality relations $\varepsilon \to \mu, \mu \to \varepsilon, E \to H, H \to -E$.

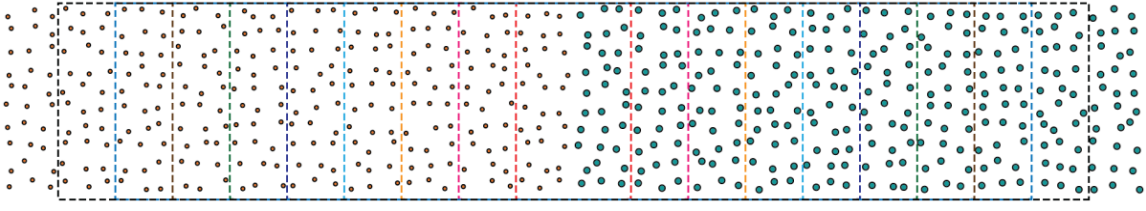

**Fig. S2. Boundary stress evaluation by Maxwell stress tensor method in the amorphous-lattice system.** Due to the random distribution of cylinders, the EM forces acting on the cylinders do not vanish immediately outside the boundary region as in the square/triangle case under the impedance-match condition. Hence the range of the boundary cannot be determined in the same way as in the square/triangle case. To determine the boundary stress, we integrate the Maxwell stress tensor over several closed paths (labelled in different colors in the figure) and then take the average $F_{boundary} = 1/J \sum_{i=1}^{J} F_i$, where $J$ is the number of integral paths considered. As long as $J$ is large enough, the fluctuation due to randomness can be averaged out and the boundary stress is given by $\sigma_x = F_{boundary}/l$ ($l$ is the width of the sample along $y$ direction). For each sample of the amorphous lattice we can calculate the corresponding boundary stress in this way and the ensemble-averaged result is taken to be the final result.